\documentclass[letterpaper]{article} 
\usepackage{aaai25}  
\usepackage{times}  
\usepackage{helvet}  
\usepackage{courier}  
\usepackage[hyphens]{url}  
\usepackage{graphicx} 
\usepackage{natbib}  
\usepackage{caption} 
\usepackage{algorithm}
\usepackage{algorithmic}

\usepackage{newfloat}
\usepackage{listings}
\usepackage{amsmath}
\usepackage{amsfonts}
\usepackage{multirow} 
\usepackage{graphicx} 
\usepackage{booktabs} 

\DeclareCaptionStyle{ruled}{labelfont=normalfont,labelsep=colon,strut=off} 
\floatstyle{ruled}
\newfloat{listing}{tb}{lst}{}
\floatname{listing}{Listing}
\title{PointTalk: Audio-Driven Dynamic Lip Point Cloud for 3D Gaussian-based Talking Head Synthesis}
\author{
    Yifan Xie\textsuperscript{\rm 1,2}, Tao Feng\textsuperscript{\rm 1}, Xin Zhang\textsuperscript{\rm 1}, Xiangyang Luo\textsuperscript{\rm 1}, Zixuan Guo\textsuperscript{\rm 3}, Weijiang Yu\textsuperscript{\rm 4}, Heng Chang\textsuperscript{\rm 5}, Fei Ma\textsuperscript{\rm 1}\thanks{Corresponding author.}, Fei Richard Yu\textsuperscript{\rm 6,7}
   \\
}
\affiliations{
    \textsuperscript{\rm 1}Guangdong Laboratory of Artificial Intelligence and Digital Economy (SZ)\\
    \textsuperscript{\rm 2}Xi'an Jiaotong University\\
    \textsuperscript{\rm 3}Peking University\\
    \textsuperscript{\rm 4}Sun Yat-sen University\\
    \textsuperscript{\rm 5}Tsinghua University\\
    \textsuperscript{\rm 6}Shenzhen University\\
    \textsuperscript{\rm 7}Carleton University\\

    \{ivanxie416, fengtao0127, zhangx0526, goodluoxy, gzx2019, weijiangyu8\}@gmail.com,
    changh17@tsinghua.org.cn,
    mafei@gml.ac.cn, richard.yu@ieee.org
}

\usepackage{bibentry}

\begin{document}

\maketitle

\begin{abstract}
Talking head synthesis with arbitrary speech audio is a crucial challenge in the field of digital humans. Recently, methods based on radiance fields have received increasing attention due to their ability to synthesize high-fidelity and identity-consistent talking heads from just a few minutes of training video. 
However, due to the limited scale of the training data, these methods often exhibit poor performance in audio-lip synchronization and visual quality.
In this paper, we propose a novel 3D Gaussian-based method called PointTalk, which constructs a static 3D Gaussian field of the head and deforms it in sync with the audio. 
It also incorporates an audio-driven dynamic lip point cloud as a critical component of the conditional information, thereby facilitating the effective synthesis of talking heads.
Specifically, the initial step involves generating the corresponding lip point cloud from the audio signal and capturing its topological structure.
The design of the dynamic difference encoder aims to capture the subtle nuances inherent in dynamic lip movements more effectively.
Furthermore, we integrate the audio-point enhancement module, which not only ensures the synchronization of the audio signal with the corresponding lip point cloud within the feature space, but also facilitates a deeper understanding of the interrelations among cross-modal conditional features.
Extensive experiments demonstrate that our method achieves superior high-fidelity and audio-lip synchronization in talking head synthesis compared to previous methods.
\end{abstract}

%

\section{Introduction}
With the development of the audio-visual industry, audio-driven talking head synthesis has become crucial in computer vision and multimedia. The goal is to create videos where a person's face moves naturally in sync with input audio while maintaining their visual identity.
This task has matured into a prominent research topic recently and holds potential for integration into diverse applications, including virtual avatars~\cite{thies2020neural}, film making~\cite{zhang2021flow}, and online meetings~\cite{kim2018deep}.
Researchers have investigated various methods to tackle this task. 
While various 2D-based studies~\cite{prajwal2020lip,wang2023seeing,zhang2023dinet,zhang2023sadtalker} successfully synthesize talking head video using generative models, the absence of a unified facial model results in shortcomings in identity preservation and pose control.
Other methods~\cite{ji2021audio,xing2023codetalker} model an individual by utilizing explicit facial structural priors, including landmarks and meshes. However, accumulated errors in these intermediate representations can significantly impact the final outcomes.

In existing research, Neural Radiance Fields (NeRF)~\cite{mildenhall2021nerf} have become popular for their ability to render realistic and 3D-consistent images from novel viewpoints, playing a crucial role in synthesizing talking head videos. 
AD-NeRF~\cite{guo2021ad} pioneered the use of conventional audio processing techniques to generate conditional features for neural radiance fields, though it faced challenges with slow training and inference speeds. 
In contrast, RAD-NeRF~\cite{tang2022real} and ER-NeRF~\cite{li2023efficient} introduced architectural innovations to streamline the enhancement of conditional features, achieving real-time inference speeds. 
Simultaneously, GeneFace~\cite{ye2022geneface} and its variant~\cite{ye2023geneface++} have employed facial landmarks as conditions to improve the generalization capabilities for out-of-domain audio.
While these methods maintain identity consistency, limited training data leads to issues with lip sync, facial details, and overall stability, reducing the realism of generated talking heads.

3D Gaussian Splatting (3DGS)~\cite{kerbl20233d} has recently achieved impressive results in 3D scene reconstruction. It utilizes 3D Gaussians as discrete geometric primitives, resulting in a clear representation of the scene and optimized real-time rendering performance. Compared to NeRF, 3DGS not only significantly improves rendering efficiency and visual quality, but also its paradigm based on 3D Gaussians is also easier to control. This potential ease of control makes it feasible to intuitively manipulate facial movements. One intuitive method is to drive a Gaussian point cloud for facial motion using parametric 3D facial models~\cite{wang2023gaussianhead,qian2024gaussianavatars}. By binding the Gaussians to the model's geometric topology, dynamic talking heads can be generated by synchronizing the displacement of the Gaussians with changes in the audio parameters.

In this paper, we propose PointTalk, a novel 3D Gaussian-based method, that attempts to utilize the 3DGS to achieve realistic and effective talking head synthesis.
Unlike previous methods that directly utilize audio signals or landmarks as conditions, 
PointTalk uses audio to generate dynamic lip point clouds, which work together with the audio signal to create more effective talking head.
Specifically, the first step involves generating a dynamic lip point cloud based on the audio signal, which captures its topological structure through a multi-resolution hash grid. 
Considering the dynamic nature of the whole process, simply capturing the features of each point cloud frame is insufficient for effectively guiding dynamic scenes. As a result, we develop a dynamic difference encoder to more accurately capture the nuances of lip movement.
Furthermore, the compression of high-dimensional conditions into significantly lower dimensions, as observed in HyperNeRF~\cite{park2021hypernerf}, results in a substantial loss of informational content. 
Therefore, we propose an audio-point enhancement module that synchronizes audio signals with point clouds and understands cross-modal feature correlations. The enhanced features produce optimal rendering results, generating high-quality syntheses. Experiments show that PointTalk renders visually realistic talking heads with accurate audio-lip synchronization.

In summary, the main contributions of our work are as follows:
\begin{itemize}
  \item We propose a novel 3D Gaussian-based framework called PointTalk, which incorporates an audio-driven dynamic lip point cloud as an additional condition to achieve realistic talking head synthesis.
  \item We introduce an Audio2Point module for generating a dynamic lip point cloud from speech audio. Additionally, we utilize a dynamic difference encoder to more accurately capture the nuances of lip movement.
  \item We design an audio-point enhancement module that synchronizes audio signals with their corresponding lip point clouds and comprehends the correlation between cross-modal conditional features.
\end{itemize}

\section{Related Work}
\subsection{Talking Head Synthesis}
Talking head synthesis aims to create a video of a speaking person that accurately represents their identity and is perfectly synchronized with the driven audio.
It consists of two main directions: 2D-based and 3D-based methods. 

\noindent \textbf{2D-Based Talking Head Synthesis.}
Some methods~\cite{prajwal2020lip,wang2023seeing,zhong2023identity,zhang2023dinet} rely on images that focus primarily on the face, particularly the mouth, to ensure the audio matches the lip movements. 
For example, Wav2Lip~\cite{prajwal2020lip} incorporates a lip synchronization expert to oversee the accuracy of lip movements, while TalkLip~\cite{wang2023seeing} utilizes a lip-reading expert to enhance the clarity and precision of these movements.
However, since these methods reconstruct the lips using only a few reference frames, they struggle to maintain consistent identity.
Recently, diffusion models~\cite{ho2020denoising} have been employed to enhance lip-sync and image quality~\cite{shen2023difftalk,ma2023dreamtalk}, but they tend to be slow during inference.
Additionally, there are some other methods~\cite{ji2022eamm,zhang2023sadtalker,ye2024real3d} that require only a single image to generate a dynamic talking head video.
For instance, 
EAMM~\cite{ji2022eamm} generates realistic emotional talking faces from a single shot.
Real3D-Portrait~\cite{ye2024real3d} enhances one-shot 3D avatar reconstruction and talking face animation.
However, such methods make it difficult to generate natural head poses and facial expressions, resulting in unrealistic visual representations.

\noindent \textbf{3D-Based Talking Head Synthesis.}
Conventional 3D-based methods~\cite{suwajanakorn2017synthesizing,ji2021audio} often utilize 3D Morphable Models (3DMM)~\cite{blanz2023morphable} for talking head synthesis. 
However, the use of intermediate representations can lead to the accumulation of errors.
With the recent rise of Neural Radiance Fields (NeRF)~\cite{mildenhall2021nerf},
it has been applied to tackle 3D head structure problems in audio-driven talking head synthesis~\cite{guo2021ad,tang2022real,li2023efficient,ye2023geneface++,shen2023sd,peng2023synctalk}.
AD-NeRF~\cite{guo2021ad} stands as the pioneering method in utilizing NeRF for audio-driven talking head synthesis.
By incorporating Instant-NGP~\cite{muller2022instant}, RAD-NeRF~\cite{tang2022real} has achieved significant enhancements in both visual quality and efficiency.
ER-NeRF~\cite{li2023efficient} introduces a triple-plane hash encoder designed to eliminate empty spatial regions and generates region-aware conditional features through an attention mechanism. 
GeneFace~\cite{ye2022geneface} and its variant~\cite{ye2023geneface++} generate content conditioned on estimated facial landmarks.
SyncTalk~\cite{peng2023synctalk} advances the realism of audio-driven talking head videos by accurately synchronizing facial identity, lip movements, expressions, and head poses. Despite its excellent performance, the inference process is not real-time.
TalkingGaussian~\cite{li2024talkinggaussian} first attempts to utilize the 3DGS~\cite{kerbl20233d} to address the facial distortion problem in existing radiance-fields-based methods. 
This paper introduces a 3D Gaussian-based method that significantly enhances visual quality and audio-lip synchronization. Moreover, it ensures that the inference process operates in real time.

\begin{figure*}[t]
\centering
\includegraphics[width=0.9\textwidth]{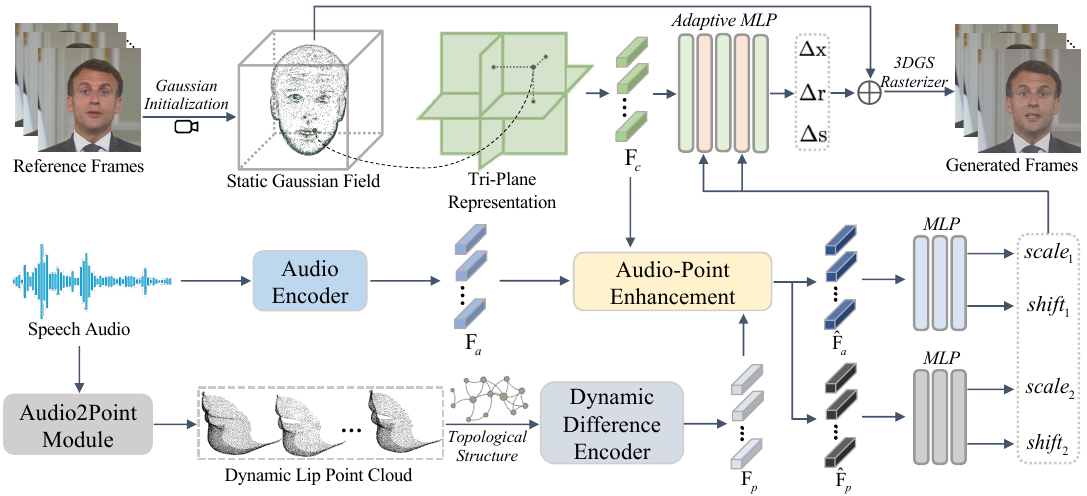} 
\caption{Overview of PointTalk. 
Utilizing the static Gaussian field to optimize the coarse Gaussian head from a random point cloud. Then the tri-plane encoder and audio encoder independently extract the spatial geometry feature $\mathrm{F}_c$ and the audio feature $\mathrm{F}_a$.
The Audio2Point module generates a dynamic lip point cloud based on the input audio signals.
Subsequently, the dynamic lip point cloud's topological structure is established, and the dynamic difference encoder extracts the point cloud feature $\mathrm{F}_p$.
Moreover, the audio-point enhancement module synchronizes the audio signals with the point cloud to facilitate information interaction, thereby obtaining the enhancement features $\mathrm{\hat{F}}_{a}$ and $\mathrm{\hat{F}}_{p}$.
Ultimately, the enhancement features are fed into two MLP decoders to compute the scale and shift factors.
By integrating these factors with $\mathrm{F}_c$, an adaptive MLP is deployed to predict the deformation parameters for 3DGS rasterizer.}
\label{figure1}
\end{figure*}

\subsection{Point Cloud Learning}
Point cloud learning~\cite{qi2017pointnet,wang2019dynamic,xie2023cross,xie2024hecpg} is a crucial research area in 3D computer vision.
PointNet~\cite{qi2017pointnet} utilizes pointwise MLPs and pooling layers to aggregate features for understanding 3D scenes.
PointNet++~\cite{qi2017pointnet++} advances PointNet~\cite{qi2017pointnet} by incorporating hierarchical sampling strategies.
Point-BERT~\cite{yu2022point} adapts the masked language modeling approach of BERT~\cite{kenton2019bert} to the 3D realm.
RECON~\cite{qi2023contrast} combines the strengths of generative and contrastive learning paradigms for enhanced 3D representation learning.
PointGPT~\cite{chen2024pointgpt} extends the concept of GPT to point clouds.
Our method generates a dynamic lip point cloud from audio and captures its topological structure using a multi-resolution hash grid.

\section{Method}
In this section, we introduce the proposed PointTalk, as illustrated in Figure~\ref{figure1}. PointTalk is structured around three key components: 1) The \textit{Multi-Attribute Branches} are utilized to process 3D Gaussians, audio signals and dynamic lip point clouds separately. 2) The \textit{Audio-Point Enhancement} module, which not only synchronizes the audio signals with the point cloud but also enhances information interaction. 3) The \textit{Adaptive 3DGS Rendering} can render the final talking head video effectively. We will delve into the specifics of these components and the associated loss functions in the subsequent subsections.

\subsection{Multi-Attribute Branches}

\noindent \textbf{3D Gaussian Branch.}
The static Gaussian field preserve the Gaussian primitives with the canonical parameters. Specifically, a Gaussian primitive can be described with a position $\mathrm{x}\in\mathbb{R}^3$, a rotation quaternion $\mathrm{r}\in\mathbb{R}^4$, a scaling factor $\mathrm{s}\in\mathbb{R}^3$, an opacity value $\alpha\in\mathbb{R}^1$, and a $d$-dimensional color feature $f\in\mathbb{R}^d$.
Consequently, the $i$-th Gaussian primitive $\mathcal{G}_i$ keeps a set of parameters $\theta_i = \{\mathrm{x}_i, \mathrm{r}_i, \mathrm{s}_i, \alpha_i, f_i\}$ and the formulation can be described as:
\begin{equation}
\mathcal{G}_{i}(p)=e^{-\frac{1}{2}(p-\mathrm{x}_i)^{T}\Sigma_{{i}}^{-1}(p-\mathrm{x}_i)},
\end{equation}
where the covariance matrix $\Sigma$ can be obtained using $\mathrm{r}$ and $\mathrm{s}$. 
We first initialize the talking head with the static Gaussian field by the training video frames to get a coarse Gaussian head. Then, we learn the deformation parameters to deform the Gaussian head with the audio.

Although Gaussian primitives are effective in representing the Gaussian head, they lack a regional position encoding due to their explicit structure. To address this, we propose utilizing a tri-plane representation~\cite{chan2022efficient}.
Specifically, for a given position $\mathrm{x}=(x,y,z)\in\mathbb{R}^{\mathrm{XYZ}}$, it is transformed through an encoding process where its projected values are utilized by three distinct 2D multi-resolution hash encoders~\cite{muller2022instant}:
\begin{equation}
\label{eq:tri_hash}
   \mathcal{H}^{\mathrm{AB}}:(a,b)\to\mathrm{f}_{ab}^{\mathrm{AB}},
\end{equation}
where the output $\mathrm{f}_{ab}^{\mathrm{AB}}\in\mathbb{R}^{LD}$, defined by $L$ levels and $D$ feature dimensions per entry, represents the planar geometric feature associated with the projected coordinate $(a, b)$. In this scenario, $\mathcal{H}^{\mathrm{AB}}$ refers to the multi-resolution hash encoder for plane $\mathbb{R}^{\mathrm{AB}}$. By concatenating these outputs, we obtain the final spatial geometry feature $\mathrm{f}_c\in\mathbb{R}^{3LD}$, which integrates the encoded geometric information. The whole process can be described as:
\begin{equation}
   \mathrm{f}_c=\mathcal{H}^\mathrm{XY}(x,y)\oplus\mathcal{H}^\mathrm{YZ}(y,z)\oplus\mathcal{H}^\mathrm{XZ}(x,z),
\end{equation}
where $\oplus$ denotes the concatenation operator.

\noindent \textbf{Audio Branch.}
In audio branch, we use an automatic speech recognition (ASR) module~\cite{amodei2016deep,hsu2021hubert} to extract audio features $\mathrm{F}_a\in\mathbb{R}^{T \times F}$ from the audio track, where $T$ denotes the frame count and $F$ represents the feature dimension. 
For additional details, please refer to the supplementary materials.

\begin{figure}[t]
\begin{center}
\includegraphics[width=0.95\linewidth]{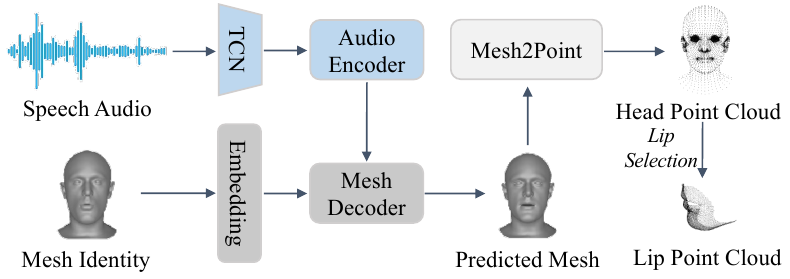}
\end{center}
\caption{The pipeline of the Audio2Point module.}
\label{figure2}
\end{figure}

\noindent \textbf{Lip Point Cloud Branch.}
Inspired by mesh-based methods~\cite{fan2022faceformer,peng2023selftalk}, we design an Audio2Point module to capture an additional dynamic lip point cloud. The pipeline of the Audio2Point module is depicted in Figure~\ref{figure2}.
Specifically, the mesh is used as the reference embedding, while the speech audio is compressed by a temporal convolutional network (TCN) and an audio encoder. Then, the predicted mesh is generated by combining the audio features with the embedding. After that, the vertices of the mesh are extracted to generate the head point cloud, which is further refined to produce the final lip point cloud ${P}\in\mathbb{R}^{T\times N\times 3}$, where $T$ represents the frame count and $N$ signifies the quantity of points.
For additional details, please refer to the supplementary materials.

For each frame of the point cloud, it is essential to capture its topological structure to achieve a robust representation. To this end, we employ $E_{\mathrm{point}}^{3}$, a multi-resolution hash grid~\cite{muller2022instant} that encodes the point cloud efficiently. 
We opt for a hash grid over a tri-plane representation when processing the lip point cloud for a key reason: the Gaussian primitives usually reach up to tens of thousands, in contrast to the lip point cloud points, which are typically in the mere hundreds. This disparity significantly reduces the potential for hash collisions.
The advantages of this way will be illustrated in subsequent ablation studies.

Given the dynamic nature of the synthesis process, we believe that merely capturing the features of each point cloud frame falls short in effectively guiding the generation of the talking head. Consequently, we design a dynamic difference encoder to better capture the nuances of lip movement. 
Specifically, the differences are taken for the features of neighboring frames, and then all the differences are concatenated together to obtain the final point cloud features $\mathrm{F}_p\in\mathbb{R}^{(T-1) \times F}$: 
\begin{equation}
\mathrm{F}_p = \mathrm{MLP}(\mathrm{cat}\left[{{E_{\mathrm{point}}^{3}(P_{t+1})}} - {{E_{\mathrm{point}}^{3}(P_{t})}}\right]_{1}^{T-1}),
\end{equation}
where $\mathrm{cat} [\cdot]$ denotes the feature concatenation and $t$ represents the frame order.

\begin{table*}[t]
\resizebox{1\linewidth}{!}{
        \setlength{\tabcolsep}{3.7mm}
        \centering
        \begin{tabular}{lcccccccc}
        \toprule
        Methods & PSNR $\uparrow$ & LPIPS $\downarrow$ & FID $\downarrow$ & LMD $\downarrow$ & LSE-D $\downarrow$ & LSE-C $\uparrow$ & Time &  FPS  \\
        Ground Truth  & N/A            & 0               & 0              & 0            & 6.897             & 8.275          & -   & -         \\ \midrule
        Wav2Lip \cite{prajwal2020lip}      & -  & -            & 11.802          & 4.498  & \textbf{7.326}      & \textbf{8.363}  & -   & 15   \\
        TalkLip \cite{wang2023seeing}       & -          & -          & 13.574         & 6.149    & 7.641      & 7.027                 & -   & 10  \\
        DINet \cite{zhang2023dinet}       & -          & -          & 7.597         & 5.712    & 8.572      & 6.358                 & -   & 17  \\
        \midrule
        AD-NeRF \cite{guo2021ad}      & 29.196          &0.1241          & 16.274          & 3.680   & 8.842       & 5.877          & 20h          & 0.1      \\
        RAD-NeRF \cite{tang2022real}     & 31.904    & 0.0722          & 9.651          & 3.197    & 8.179      & 6.043                & 6h  & {28}        \\ 
        GeneFace++ \cite{ye2023geneface++}& 30.417    & 0.0968         & 11.709          & 3.546    & 7.547      & 6.335             & 8h  & 23  \\
         ER-NeRF \cite{li2023efficient}     & 31.959   & 0.0379         & \textbf{6.927}          & 3.125     & 8.254     & 6.174         & \underline{2h}  & {30}   \\
         TalkingGaussian \cite{li2024talkinggaussian}     & \underline{32.398}   & \underline{0.0355}         & 8.385          & \underline{2.967}     & 7.825     & 6.516         & \textbf{1h}  & \textbf{90}   \\
        \midrule
        PointTalk & \textbf{32.770}          & \textbf{0.0337} & \underline{7.331} & \textbf{2.818} & \underline{7.383}& \underline{7.165}        & \textbf{1h}  &\underline{85}   \\ \bottomrule 
        \end{tabular}
    }
    \setlength{\abovecaptionskip}{0.25cm}
    \caption{\textbf{The quantitative results of the head reconstruction}. The boldface indicates the best performance and the underline represents the second-best performance. 
    In the self-driven evaluation, since Wav2Lip, TalkLip, and DINet have access to the ground truth with the exception of the mouth region, the PSNR and LPIPS metrics are deemed inapplicable.
    }
    \label{tab:setting1}
    \vspace{-0.5cm}
\end{table*}

\subsection{Audio-Point Enhancement}
\begin{figure}[t]
\begin{center}
\includegraphics[width=0.9\linewidth]{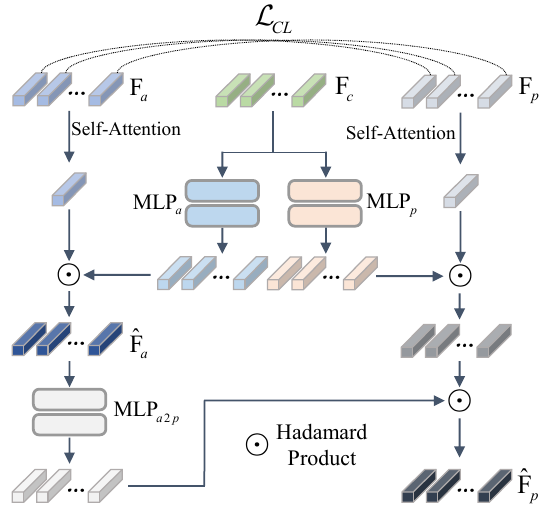}
\end{center}
\caption{The detailed structure of the Audio-Point Enhancement module.}
\label{figure3}
\vskip -10pt
\end{figure}

We introduce the audio-point enhancement module, designed with a dual purpose: to synchronize the audio signals with the corresponding point cloud, and to understand the correlation between cross-modal features. The detailed structure of this module is depicted in Figure~\ref{figure3}.

Drawing inspiration from~\cite{chung2017out,prajwal2020lip}, we recognize that synchronizing the audio signals with the video signals can facilitate the generation of the talking head to a notable extent. In our method, we assert that synchronizing the audio signals with the corresponding point cloud will likewise contribute to enhanced performance.
Specifically, we introduce a cross-modal contrastive learning strategy to establish the audio-point correspondences.

Given the audio features $\mathrm{F}_a\in\mathbb{R}^{T \times F}$ and point cloud frame features $\mathrm{F}'_p \in\mathbb{R}^{T \times F}$, 
the process of individually extracting lip point cloud features for each frame is aimed at establishing a correspondence with the audio features.
For the $t$-th frame, our objective is to enhance the similarity between $\mathrm{F}_{at}$ and $\mathrm{F}'_{pt}={E_{\mathrm{point}}^{3}(P_{t})}$, as they both correspond to the same object. Simultaneously, we strive to reduce the similarity between $\mathrm{F}_{at}$ and the point cloud features of non-corresponding frames.
Therefore, we can construct the loss function $l(t, \mathrm{F}_{a}, \mathrm{F}'_{p})$:
\begin{equation}
    l(t, \mathrm{F}_{a}, \mathrm{F}'_{p}) = 
    -\log
    \frac{ {\exp(sim(\mathrm{F}_{at}, \mathrm{F}'_{pt})/\tau)}}
    {{\sum\limits_{\substack{k=1 \\ k \neq t}}^{T} \exp(sim({\mathrm{F}_{at}, \mathrm{F}'_{pk})/\tau)}}},
\end{equation}
where $\tau$ is the temperature factor and $sim (\cdot,\cdot)$ denotes the cosine similarity function. And the cross-modal contrastive learning loss $\mathcal{L}_{CL}$ is then formulated as:
\begin{equation}
\label{eq:lcl}
    \mathcal{L}_{{CL}}= \sum_{t=1}^{T}[ l(t, \mathrm{F}_{a}, \mathrm{F}'_{p})
    + l(t, \mathrm{F}'_{p}, \mathrm{F}_{a})].
\end{equation}

Dynamic conditions, including audio signals and lip point clouds, play a selective role in talking head synthesis. Many previous methods~\cite{guo2021ad,tang2022real} have simply assumed that these dynamic conditions uniformly influence the entire synthesis process.
To comprehend the correlation between cross-modal features, lightweight external attention~\cite{guo2022beyond, li2023efficient} is employed for information interactions. Specifically, self-attention is first utilized to compress audio and point cloud features following ~\cite{tang2022real}. A two-layer MLP is then utilized to capture the global context of spatial geometry. Following this, we apply the Hadamard product to multiply the global context with the compressed dynamic conditions, yielding the final enhanced features. This process can be represented as follows:
\begin{equation}
\mathrm{\hat{F}}_{a} = 
\mathrm{MLP}_a (\mathrm{F}_c) \odot \mathrm{SA} (\mathrm{F}_a),
\end{equation}
where $\odot$ denotes Hadamard product. A similar process is performed for the point cloud.

Additionally, given that the whole process is audio-driven and the lip point cloud is generated from the audio signals, we aim for the point cloud features to emphasize the audio-related parts more prominently. To achieve this, we process the enhanced audio features through another two-layer MLP to capture the global content of the audio. Subsequently, we utilize the Hadamard product to derive the final enhanced point cloud features. 

\begin{table}[]
\resizebox{\linewidth}{!}{
\begin{tabular}{@{}lcccc@{}}
\toprule
 & \multicolumn{2}{c}{Audio A}       & \multicolumn{2}{c}{Audio B}       \\ \cmidrule(l){2-5} 
Methods               & LSE-D ↓         & LSE-C ↑         & LSE-D ↓         & LSE-C ↑         \\ 
Ground Truth & 6.899           & 7.354         & 7.322             & 8.682         \\
\midrule

Wav2Lip \cite{prajwal2020lip}                                         & \textbf{7.896}          & \textbf{7.393}          & \textbf{6.760}          & \textbf{9.259}         \\
TalkLip \cite{wang2023seeing}                                      & 9.013           & 6.229         & 8.705          & \underline{7.528}          \\
DINet \cite{zhang2023dinet}                                      & 8.771          & 6.196          & 8.746        & 7.140        \\
\midrule
AD-NeRF \cite{guo2021ad}                                      & 14.432          & 1.274          & 13.896         & 1.877        \\
RAD-NeRF \cite{tang2022real}                                       & 11.639         & 1.941          & 11.082          & 3.135           \\
GeneFace++ \cite{ye2023geneface++}                                    & 9.195         & 4.868          & 8.540          & 6.631          \\
ER-NeRF \cite{li2023efficient}                                       & 9.569           &  4.670        & 9.082        & 5.871        \\
TalkingGaussian \cite{li2024talkinggaussian}
&9.171            &  5.327        & 9.061        & 5.745        \\
\midrule
PointTalk  & \underline{8.406}    & \underline{6.427}    & \underline{8.331} & 7.018 \\ \bottomrule
\end{tabular}
}
\caption{\textbf{The quantitative results of the lip synchronization}. We utilize two different audio samples to driven the same subject. The boldface indicates the best performance and the underline represents the second-best performance.}
\label{tab:setting2}
\end{table}

\begin{figure*}
  \centering
  \includegraphics[width=0.9\linewidth]{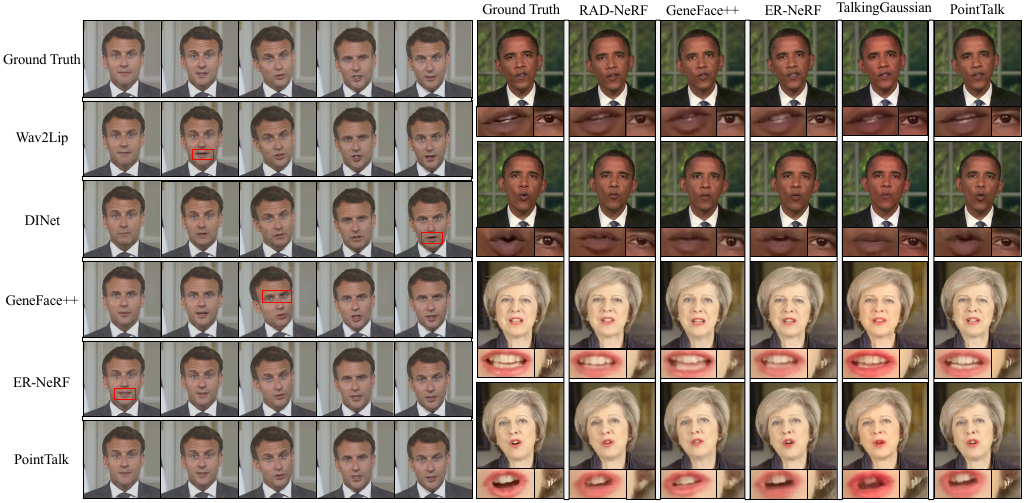}
  \caption{Qualitative comparison of talking head synthesis by different methods. PointTalk has the best visual effect on lip movements and facial details. Please zoom in for better visualization.
  }
  \label{figure4}
\end{figure*}

\subsection{Adaptive 3DGS Rendering}
Most previous methods~\cite{guo2021ad,tang2022real,li2023efficient} typically integrate conditional features into spatial geometry features through concatenation. 
However, we argue that concatenation alone is insufficient for the effective fusion of cross-modal features.
Inspired by Adaptive Instance Normalization (AdaIN)~\cite{huang2017arbitrary}, we adopt a combination of a residual block and AdaIN to guide the fusion process.
This process subjects the outputs to an affine transformation, incorporating both translation and scaling, parameterized by $scale_i$ and $shift_i$, respectively. These parameters are then employed to generate adaptive features $\mathrm{F}'_c$:
\begin{equation}
\mathrm{F}'_c = (1 + scale_i)*\mathrm{F}_c + shift_i.
\end{equation}
Ultimately, we can use the adaptive features to derive the point-wise deformation parameters ($\Delta \mathrm{x}, \Delta \mathrm{r}, \Delta \mathrm{s}$) for the 3DGS Rasterizer, which are irrelevant to the color and opacity changes.
Therefore, we can calculate the final parameter $\theta_D = \{\mathrm{x}+\Delta \mathrm{x}, \mathrm{r}+\Delta \mathrm{r}, \mathrm{s}+\Delta \mathrm{s}, \alpha, f\}$ for rendering.
The 3DGS rasterizer gather $N$ modified Gaussians with the camera information to compute the color $\mathcal{C}$ of pixel $p$:
\begin{equation}
\mathcal{C}(p)=\sum_{i\in N}c_{i}\hat{\alpha}_{i}\prod_{j=1}^{i-1}(1-\hat{\alpha}_{j}),
\end{equation}
where $c_i$ represents the decoded color from $f$, $\hat{\alpha_i}$ is the result of calculating opacity $\alpha_i$ with the projected function. 
For a more comprehensive understanding of the 3DGS Rasterizer, please refer to the supplementary materials.

\subsection{Loss Function}
To initialize the coarse Gaussian head, we follow the original 3DGS~\cite{kerbl20233d} and utilize a combination of pixel-wise loss $\mathcal{L}_1$ and D-SSIM term $\mathcal{L}_{D-SSIM}$ (weighted by $\lambda_1$).
After the initialization, we further predict the deformation parameters and input them for the 3DGS rasterizer to render the output images. 
Following the process of previous methods~\cite{tang2022real,li2023efficient}, we randomly sample a set of patches from the entire image and incorporate LPIPS loss~\cite{zhang2018unreasonable} (weighted by $\lambda_2$) to improve detail resolution.
This is combined with the cross-modal contrastive learning loss $\mathcal{L}_{{CL}}$ (weighted by $\lambda_3$) as described in Eq.~\ref{eq:lcl}. The overall loss function can be construsted as:
\begin{equation}
\begin{aligned}
\mathcal{L}=\mathcal{L}_{1}+\lambda_1\mathcal{L}_{{D-SSIM}}+\lambda_2\mathcal{L}_{{LPIPS}}+\lambda_3\mathcal{L}_{{CL}},
\end{aligned}
\end{equation}

\section{Experiments}

\subsection{Experimental Settings}

\noindent \textbf{Dataset.}
To ensure a fair comparison, the dataset for our experiments is sourced from ~\cite{tang2022real,ye2022geneface, li2023efficient} and includes both English and French languages. 
We collect high-definition speaking video clips, each with an average length of approximately 7,500 frames at 25 FPS. 
Each raw video is cropped and resized to a resolution of 512$\times$512, focusing on a centered portrait.

\noindent \textbf{Comparison Baselines.}
We compare our method against three 2D-based methods, such as Wav2Lip~\cite{prajwal2020lip}, TalkLip~\cite{wang2023seeing}, and DINet~\cite{zhang2023dinet}, as well as five 3D-based methods, including AD-NeRF~\cite{guo2021ad}, RAD-NeRF~\cite{tang2022real}, GeneFace++~\cite{ye2023geneface++}, ER-NeRF~\cite{li2023efficient}, and TalkingGaussian~\cite{li2024talkinggaussian}.

\subsection{Quantitative Evaluation}
\noindent \textbf{Metrics.}
We utilize Peak Signal-to-Noise Ratio (PSNR) to assess the overall image quality and Learned Perceptual Image Patch Similarity (LPIPS)~\cite{zhang2018unreasonable} to evaluate the details. 
Additionally, we employ \text{Fréchet} Inception Distance (FID)~\cite{heusel2017gans} to gauge image quality at the feature level. 
For evaluating lip synchronization, we recommend using the Landmark Distance (LMD), which quantifies the distance between lip landmarks. 
Furthermore, we introduce Lip Sync Error Distance (LSE-D) and Lip Sync Error Confidence (LSE-C), consistent with Wav2Lip~\cite{prajwal2020lip}, to assess the synchronization between lip movements and audio.

\noindent \textbf{Comparison Settings.}
In our quantitative evaluation, we assess our method in two distinct settings: the head reconstruction setting and the lip synchronization setting. 
For the head reconstruction setting, we divide each video into training and test datasets to evaluate the quality of the talking head reconstruction. 
For the lip synchronization setting, we extract two out-of-distribution audio clips, named Audio A and Audio B.
These audio clips are utilized to drive the same subject for comparison in lip synchronization.

\noindent \textbf{Evaluation Results.}
The evaluation results of the head reconstruction setting are illustrated in Table~\ref{tab:setting1}.
It can be observed that our image quality is superior to other methods in almost all aspects. In terms of lip synchronization, our results surpass most methods.
Specifically, while one-shot methods such as Wav2Lip~\cite{prajwal2020lip}, Talklip~\cite{wang2023seeing}, and DINet~\cite{zhang2023dinet} perform excellently on the LSE-D and LSE-C metrics and can synthesize talking heads without per-identity training, they score poorly on other metrics. 
Compared to other 3D-based methods, our method outperforms them in most metrics. Especially in lip synchronization metrics, the improvement of our method is more obvious. This is mainly due to the assistance of the dynamic lip point cloud.
In terms of lip synchronization, the evaluation results are shown in Table~\ref{tab:setting2}.
Our method also demonstrates an excellent generalization ability to synthesize lip-sync talking heads.
Additionally, our method maintains superior performance in both training time and inference FPS, approximating TalkingGaussian~\cite{li2024talkinggaussian}, which demonstrates the high efficiency of PointTalk.

\subsection{Qualitative Evaluation}
\noindent \textbf{Evaluation Results.}
To more intuitively evaluate image quality and lip synchronization, we present a comparison of our method with others in Figure~\ref{figure4}. 
We showcase key frames from a clip and close-up details of two talking heads. 
The results demonstrate that our PointTalk captures finer details and achieves the highest accuracy in lip synchronization. 

\begin{figure}[t]
\begin{center}
\includegraphics[width=0.95\linewidth]{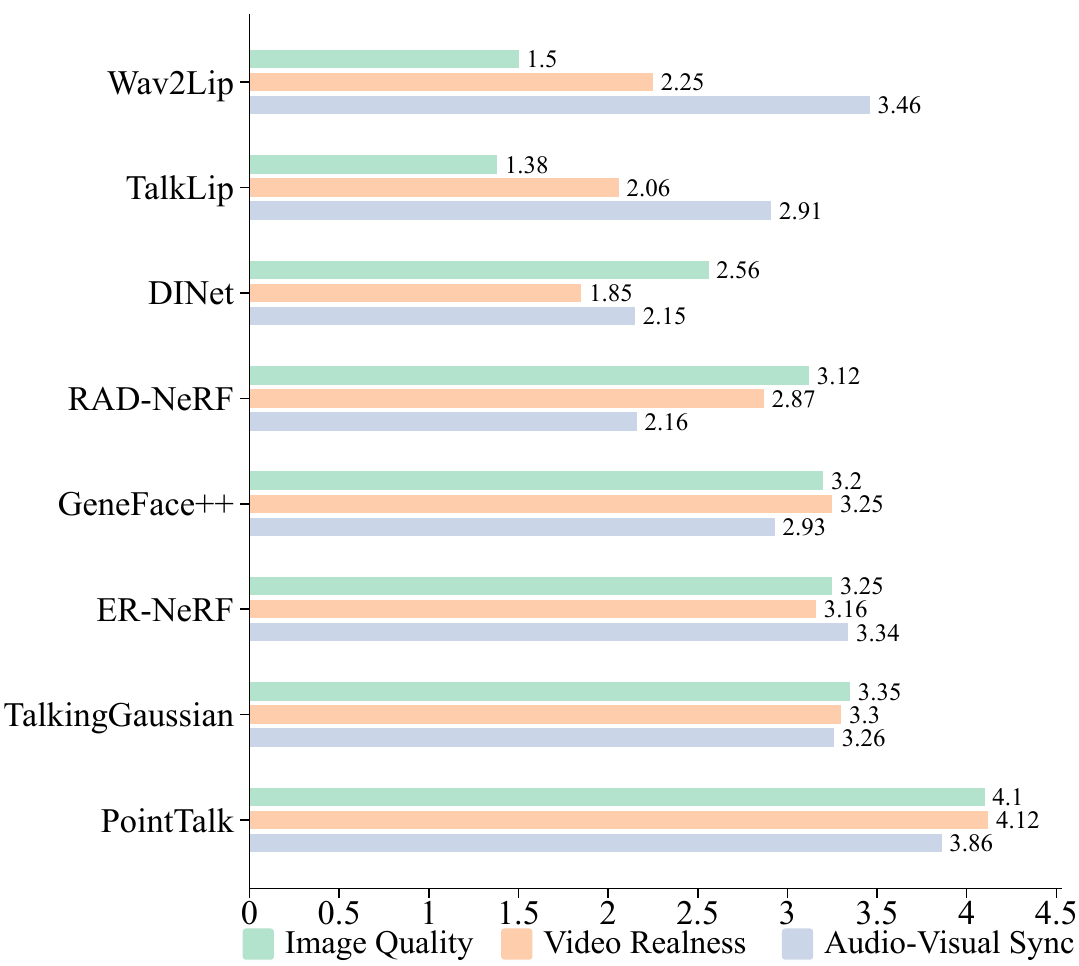}
\end{center}
\caption{User study. The rating scale ranges from 1 to 5, with higher numbers indicating better performance.}
\label{figure5}
\end{figure}

\begin{figure}[t]
\begin{center}
\includegraphics[width=0.9\linewidth]{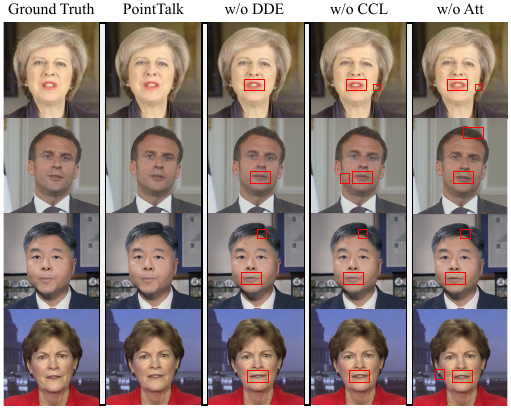}
\end{center}
\caption{Visualization of the ablation study.}
\label{figure6}
\end{figure}


\noindent \textbf{User Study.}
To conduct a more comprehensive evaluation of PointTalk, we implement a user study questionnaire. 
We select 36 video clips generated during the quantitative evaluation and invite 16 participants to take part in the survey. 
Participants are required to rate the generated videos based on three aspects: (1) Image Quality, (2) Video Realness, and (3) Audio-Visual Synchronization. 
The average scores for each criterion are presented in Figure~\ref{figure5}. 
PointTalk outperforms previous methods in all aspects. 
Notably, in terms of image quality and video realness, PointTalk exceeds the second-ranked methods by a margin of over 20\%. 

\subsection{Ablation Study}
We conduct an ablation study under the head reconstruction setting to assess the contributions of various components in our PointTalk. 
The results are presented in Table~\ref{tab:ablation}.

\noindent \textbf{Lip Point Cloud Encoder.}
In our method, we utilize a multi-resolution hash grid to capture the topological structure of the lip point cloud. This setting is replaced with a graph neural network (GNN)~\cite{wang2019dynamic} and tri-plane representation~\cite{li2023efficient}, as illustrated in Table~\ref{tab:ablation} (lines 1-2). 
The findings reveal that solely employing a GNN falls short in efficiently extracting the topological structure. Moreover, the tri-plane representation tends to induce a certain level of information loss. 
Further investigation into the optimal level ($\mathrm{L}$) and dimension ($\mathrm{F}$) of the multi-resolution hash grid is presented in Table~\ref{tab:ablation} (lines 4-7). Optimal performance is achieved at $\mathrm{L}=8$ and $\mathrm{F}=4$, which we adopt for all our experiments. 
Additionally, we develop a dynamic difference encoder (DDE) to better capture the subtle movements of the lips.
A comparison of the results in Table~\ref{tab:ablation} (lines 8 and 11) clearly shows that the DDE enhances lip-sync metrics. Figure~\ref{figure6} further demonstrates the inaccuracies in lip shape that occur without the use of DDE.

\begin{table}[t]
\centering
\resizebox{1\linewidth}{!}{
    \setlength{\tabcolsep}{1mm}
    \begin{tabular}{lcccccc}
    \toprule
    Setting  & PSNR$\uparrow$ & LPIPS$\downarrow$  & FID$\downarrow$ & LMD$\downarrow$  & LSE-D$\downarrow$ & LSE-C$\uparrow$\\ \midrule
    GNN & 31.011  & 0.0416       & 9.353      & 3.655  & 8.135  & 6.253 \\
    Tri-Plane & 32.390 & 0.0389 & 7.319 &2.873  & 7.636 & 6.816\\ 
    Hash Grid & \textbf{32.704}  & \textbf{0.0365} & \textbf{7.194} & \textbf{2.773} & \textbf{7.367} & \textbf{7.079}   \\
    \bottomrule
    L=32 F=1 & 32.515  & 0.0378       & 7.324      & 2.915 & 7.564 & 6.818    \\
    L=16 F=2 & 32.662  & 0.0367       & 7.295      & 2.864 & 7.471 & 6.995    \\
    L=8 F=4 & \textbf{32.704}  & \textbf{0.0365} & \textbf{7.194} & \textbf{2.773} & \textbf{7.367} & \textbf{7.079}   \\
    L=4 F=8 & 32.617  & 0.0376      & 7.523     & 2.818  & 7.679 & 6.790   \\
    \bottomrule
    w/o DDE & 32.032  & 0.0384& \textbf{6.997}     & 2.976 & 7.649 & 6.358    \\
    w/o CCL & 32.125  & 0.0389  & 7.357  & 2.953  & 7.786 & 6.455   \\
    w/o Att & 31.790 & 0.0418   & 8.353  & 3.115& 8.118 & 6.124    \\
    PointTalk & \textbf{32.704}  & \textbf{0.0365} & 7.194 & \textbf{2.773} & \textbf{7.367} & \textbf{7.079}    \\
    \bottomrule
    \end{tabular}
}
\caption{Ablation study on different settings. Best performance is highlighted in bold.}
\setlength{\abovecaptionskip}{0cm}
\label{tab:ablation}
\end{table}

\noindent \textbf{Audio-Point Enhancement.}
To evaluate the Audio-Point Enhancement module's effectiveness, experiments in Table~\ref{tab:ablation} (lines 9-11) show that removing cross-modal contrastive learning (CCL) causes misalignment between audio and lip points, while excluding external attention (Att) weakens cross-modal feature correlation. As shown in Figure~\ref{figure6}, these ablations result in inaccurate lip shapes and blurred details.

\section{Conclusion}
In this paper, we propose PointTalk, a novel 3D-Gaussian based method that incorporates an audio-driven dynamic lip point cloud to achieve realistic talking head synthesis. 
Initially, we introduce an Audio2Point module for generating a dynamic lip point cloud and develop a dynamic difference encoder to precisely encode the nuances of lip movement.
Furthermore, we integrate an audio-point enhancement module. This module not only synchronizes audio signals with their corresponding lip point clouds but also understands the correlation between cross-modal conditional features. 
Extensive experiments in various settings demonstrate the superior performance of PointTalk.


\end{document}